# Visible optical vortices measured with bulk lateral shearing interferometry

**MIGUEL LÓPEZ-RIPA[1], ÍÑIGO SOLA[1, 2], AND BENJAMÍN ALONSO[1,2,*]**

[1] *Grupo de Investigación en Aplicaciones del Láser y Fotónica, Departamento de Física Aplicada, Universidad de Salamanca, E-37008 Salamanca, Spain*
[2] *Unidad de Excelencia en Luz y Materia Estructuradas (LUMES), Universidad de Salamanca, E-37008 Salamanca, Spain*
**b.alonso@usal.es*

**Abstract:** Ultrafast pulse optical vortices are spatiotemporal structures with a diverse range of applications. There are different ways to generate them, often restricted to a wavelength range. Likewise, characterization techniques frequently possess similar limitations. In this work, we first generate ultrashort optical vortices in the near infrared from Ti:sapphire laser pulses by means of structured waveplates and beam manipulation. Then, we produce the visible vortices through up-conversion using a second-harmonic generation crystal. The resulting beams require spatiotemporal characterization, which are performed by bulk lateral shearing interferometry. The reference pulse is temporally characterized with the amplitude swing technique. In this manner, we present the generation of these pulses in the visible range, which are experimentally validated, and demonstrate that bulk lateral shearing interferometry can be used for pulsed beams across widely different spectral regions with the same setup. This finding is significant for future applications of the technique with various sources.

## 1. Introduction

Ultrashort laser pulses are one of the shortest controllable events in nature, with time evolution in the femtosecond scale. Today, complex shaping and production of structures in their spatiotemporal and spatio-spectral dependences [1] are more and more ubiquitous in their applications [2]. From a basic point of view, spatiotemporal couplings as pulse-front tilt, spatial chirp, angular dispersion or combinations of them [3], arises from common elements as tilted windows, gratings [4], prisms, or lenses [5] or diffractive optical elements [6–8], among others. Other applications involving attosecond pulse generation [9], nonlinear propagation in filamentation [8,10], propagation in waveguides [11], laser plasma acceleration [12], and many more, show a rich phenomenology of spatiotemporal couplings too.

Optical vortex beams carrying orbital angular momentum (OAM) of light are also used in many applications [13,14], very often involving ultrashort laser pulses [15–17]. Optical vortices have been produced in different spectral ranges from the mid-infrared [18] to the X-rays [19], and with a wide variety of methods, such as spiral phase plates [20], holographic structures [21], spatial light modulators [22] or structured waveplates [14]. Their characterization requires a precise wavefront measurement capable of reconstructing the beam singularity [23] and, due to this, sometimes interferometric detections are applied to evaluating the topological charge instead of a full wavefront measurement.

The characterization of spatiotemporal couplings of pulsed beams is important for the applications [2]. Different techniques have been used to address the simultaneous determination of the space and time (or frequency) dependence of the electric field of ultrashort laser pulses [24,25], for example STARFISH [26], spatially resolved Fourier transform spectrometry [27], TERMITES [28], INSIGHT [29] or FALCON [30]. Also, the spatiotemporal evolution of arbitrarily polarized beams has already been addressed [31]. Due to the need to resolve the field in multiple dimensions, these techniques are based on different strategies more inspired by spatial detections [23,32–34] or temporal detections [35,36], which are translated into techniques having more resolution in the spatial or in the temporal dimension. Typically, the techniques are designed for a particular spectral region or have bandwidth limiting components. In some scanning techniques that can measure the wavefront, the shot-to-shot instability may introduce some noise in the measured wavefront even for partially common-path interferometers [37,38]. The technique Bulk LAteral SHearing Interferometry (BLASHI) was recently introduced [39], based on lateral shearing interferometry [32] and spectral interferometry [40]. This interferometry is implemented by means of a monolithic configuration with a uniaxial crystal and a linear polarizer that create the two interfering beams, showing a ultrastable interference because of the compact common-path setup. Previously, it has been found that such kind of bulk interferometer is not affected by air flows, vibrations or temperature variations, presenting an excellent stability of $\sim\lambda/10^3$ [41], which can be neglected compared to other standard or partially common-path interferometers [37,38,42,43]. In principle, the use of a uniaxial crystal opens the way to using the BLASHI technique anywhere within its transparency band. Here, we propose to use BLASHI to measure near-infrared (NIR) and visible (VIS) spatiotemporal structures with the same setup.

In this work, we present the spectral dependence of the parameters involved in BLASHI due to the uniaxial crystal (i.e., the walk-off and the delay or dispersion of the crystal for different spectral regions). We use structured waveplates to generate optical vortices in the NIR [44] and in the VIS through up-conversion. We apply the BLASHI technique to measure said vortices with different OAMs, while we use the amplitude swing technique [45] to measure a reference pulse in the temporal (spectral) domain. We find that BLASHI can retrieve the spatio-spectral (and, equivalently, the spatiotemporal) intensity and phase of complex beam structures with precision in the NIR and in the VIS, obtaining results consistent with the prediction for the generated beams.

## 2. Materials and Methods

### *2.1 Bulk lateral shearing interferometry*

The technique Bulk LAteral SHearing Interferometry (BLASHI) [39] combines two well-known aspects of interferometry to obtain the space-frequency (or space-time) characteristics of a beam. Firstly, it is based on lateral shearing (a lateral spatial displacement) interferometry [32,46] through the measurement of the interference of a beam with a spatially sheared replica of itself to obtain the spatially varying phase, i.e., the wavefront. Secondly, it relies on spectral interferometry of two delayed pulses [40] to obtain the time (and frequency) evolution of the beam. BLASHI allows simultaneously obtaining the couplings between space and frequency features, therefore being able to reconstruct the spatiotemporal evolution of the beam.

The bulk implementation of the technique consists of using a plane-parallel uniaxial crystal with its optical axis (OA) oblique to the plate faces (see Figure 1). Due to the birefringence, light propagating through the crystal is separated into two beams with orthogonal linear polarization, the so-called ordinary and extraordinary beams. For an input beam with linear polarization with the appropriate orientation, said beams are created with the same energy. Under normal incidence, at the output of the crystal the two beams propagate parallel but with a lateral shear due to the walk-off $\Delta$ and with a temporal delay $\tau$ due to the birefringence. In order to observe their interference, a linear polarizer is used to project the beams into the same direction. The polarizer orientation is set at an intermediate orientation to have the same energy balance, thus having nearly maximum contrast. The interference signal is recorded by spatially scanning an optical fiber connected to a commercial spectrometer (HR4000 Ocean Optics Inc. for the VIS, Avantes AvaSpec-2048 for the NIR).

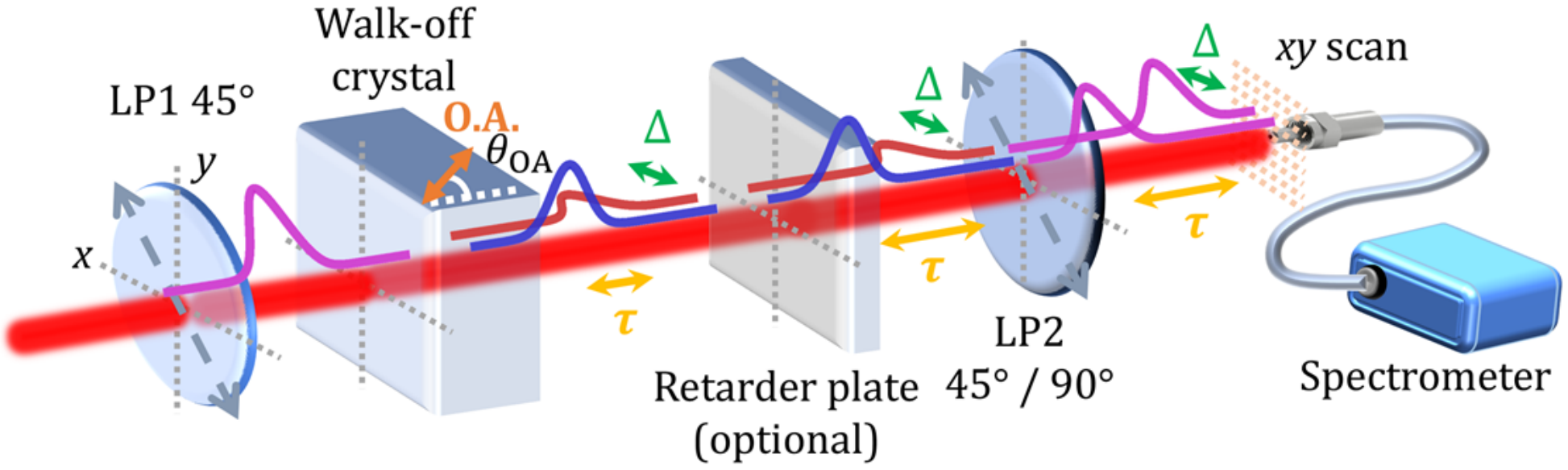

Figure 1. Diagram of the BLASHI method for the case with walk-off in the $x$-axis. A uniaxial (walk-off) crystal (optical axis, O.A.) is used to divide the input beam (purple, linearly polarized at 45°) into two: the ordinary (blue) and the extraordinary (red) beam, which are linearly polarized at 90° (vertical) and 0° (horizontal), respectively. Both beams get a relative delay ($\tau$), and the extraordinary beam acquires a lateral shear due to the walk-off ($\Delta$). An optional retarder plate can adjust the delay if needed. A linear polarizer projects both beams into the desired polarization (45° to measure the interferences and 90° to measure the spectral amplitude of the ordinary beam). The spectral signal is recorded by raster scanning in 2D ($x$,$y$) a fiber spectrometer.

For a plane-parallel uniaxial crystal with the optical axis oriented at an angle $\theta_{OA}$ with respect to the normal of the input face, the walk-off imparted to the input beam under normal incidence will be produced in the direction corresponding to the projection of the optical axis onto the face of the crystal. The walk-off $\Delta$ (i.e., the lateral shear) is proportional to the crystal thickness $L$ and depends on the angle $\theta_{OA}$ and its ordinary and extraordinary refractive indices, $n_o$ and $n_e$ respectively. Solving the beam propagation, it can be calculated as

$$\Delta = \frac{L\sin(2\theta_{OA})}{2}\frac{n_e^2 - n_o^2}{n_e^2\cos^2\theta_{OA} + n_o^2\sin^2\theta_{OA}} = \frac{L\sin(2\theta_{OA})\, n_e^2(\theta_{OA})}{2}\left(\frac{1}{n_o^2} - \frac{1}{n_e^2}\right) \tag{1}$$

where it is important to remark that for operating at different spectral ranges, the wavelength-dependent refractive indices are considered. In this equation, the notation $n_e(\theta_{OA})$ refers to the effective refractive index for an extraordinary beam propagating at an angle $\theta_{OA}$ with respect to the optical axis. In our case, the dispersion of the calcite response is calculated from the Sellmeier equations [47]. For moderate bandwidth of the pulses, the walk-off can be assumed to be nearly constant, considering the spatial resolution of 4 μm of the single-mode collecting fiber. However, when measuring beams in different spectral bands, the walk-off deviation is not negligible. In Fig. 2(a) we show the walk-off dispersion, where the values at 800 nm and 400 nm (113 μm and 123 μm, respectively) used in this work are highlighted (for the 1.06-mm-thick calcite plate with $\theta_{OA} = 45°$).

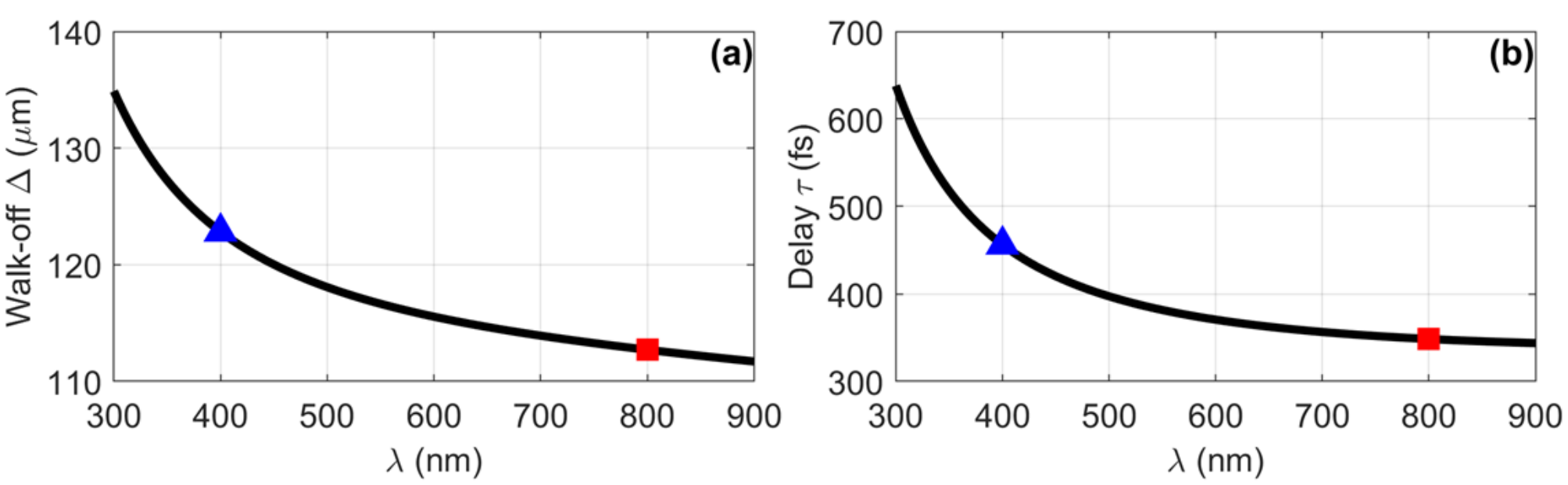

Figure 2. Spectral dependence of the (a) walk-off $\Delta$ and (b) the delay $\tau$ under normal incidence for a 1.06-mm-thick calcite with cut angle $\theta_{OA} = 45°$.

As mentioned earlier, the walk-off depends on the wavelength, and for moderate spectral bandwidths, the effect is negligible. For instance, in the crystal analyzed here, there is a walk-off variation of approximately 2 μm over a spectral range of 200 nm. However, for ultrabroadband pulses with larger bandwidths (e.g., few-cycle pulses), the effect may no longer be negligible, and alternative strategies should be considered to address it.

As said before, the birefringence is responsible for the delay $\tau$ between the ordinary and extraordinary beams, which can be calculated from the difference of phase, $\Delta\phi(\omega)$, acquired by the two beams during their propagation through the crystal

$$\tau = \frac{\partial}{\partial\omega}[\Delta\phi(\omega)] = \frac{\partial}{\partial\omega}\left[(n_o(\omega) - n_e(\theta_{OA},\omega))\frac{\omega}{c}L\right] \tag{2}$$

where c is the speed of light in vacuum..

Note that, due to calcite crystal having negative birefringence ($n_o > n_e$), the ordinary beam is delayed relative to the extraordinary beam. If the crystal has positive birefringence, it simply must be considered for the walk-off orientation and the delay sign in the interferometry. For the calcite crystal described before, the delay calculated from Eq. (2) is depicted in Fig. 2(b), where we highlight the values 348.4 fs and 456.3 fs (at 800 nm and 400 nm, respectively). Related to the delay, two points must be clarified. Firstly, the appropriate delay for spectral interferometry must be chosen within a range according to resolve the interferences and to allow Fourier-transform analysis [40]. Since the beam replicas have orthogonal polarization, the delay can be readily adjusted with a common retarder plate, in our case we used a 3-mm-thick calcite (1.8 ps delay at 800 nm) to increase the delay. Secondly, despite the relative phase $\Delta\phi(\omega)$ can be calculated from Eq. (2), we choose to measure it, so that the effect of thickness can be accurately calibrated [39]. Note that said phase difference is frequency dependent, being the effect of dispersion computed within said calibration. The dephase introduced by the optional retarder plate (to adjust the delay) can be calibrated included in the global phase of the system or separately with in-line single-channel spectral interferometry [41]. Note that the delay dispersion described by Eq. (2) is ultimately encompassed within the frequency-dependent phase difference discussed earlier. This relative phase is calibrated and then subtracted within the retrieval algorithm. Even when using ultrabroadband pulses, which will experience different chirping, this effect does not impact the reconstructed beam.

In the case of introducing the delay in the $x$-axis, due to the beam replicas being identical (except for the shear and delay), their spectral interference $S_{SI}$ recorded at different spatial positions $(x,y)$ will depend on the spectrum of the beam at said point, $S(x,y,\omega)$, and will encode the gradient of the spatio-spectral phase, $\varphi(x,y,\omega)$, in the cosine term given by the following equation

$$S_{SI}(x,y,\omega) = \frac{1}{2}S(x,y,\omega) + \frac{1}{2}S(x-\Delta,y,\omega) + \sqrt{S(x,y,\omega)S(x-\Delta,y,\omega)}\cos\left[\varphi(x,y,\omega) - \varphi(x-\Delta,y,\omega) + \Delta\phi_{\mathrm{cal}}(\omega)\right] \tag{3}$$

where $\Delta\phi_{\mathrm{cal}}(\omega)$ is the calibrated dephase between the ordinary and extraordinary beams (including the uniaxial crystal and the optional retarder plate). From Fourier analysis [40] of said interference it is possible to retrieve the $x$-gradient of $\varphi(x,y,\omega)$ so that it can be integrated in $x$ afterwards [39]. The same measurement must be taken for walk-off in the $y$-axis (just rotating 90° the walk-off crystal), so that the $y$-gradient can be encoded and integrated. The reconstructed phase this way is the frequency-resolved wavefront. The spectrum dependence $S(x,y,\omega)$ can be obtained from the crossed term of Eq. (3) assuming slow spatial variation in a step $\Delta$ or can be directly experimentally acquired by filtering the ordinary beam with the linear polarizer. The phase obtained is relative to a initial spatial position in the gradient integration. To have the full information, a reference pulse must be measured at any coordinate $(x,y)$ with one temporal (spectral) diagnostics such as FROG [35], SPIDER [48], d-scan [49] or a-swing [45]. Otherwise, the information will be relative rather than absolute but will still provide valuable details of the spatio-spectral couplings.

Finally, from the spatio-spectral amplitude and phase described above, it is straightforward to obtain the full spatiotemporal information just by Fourier transforming it.

### *2.2. Generation of optical vortices*

The optical vortices were generated in the NIR from a femtosecond laser at ~800 nm using structured waveplates (s-plates) [14,50]. These s-plates are laser inscribed nanogratings in glass with tailored birefringence (dephase and axes orientation), which can be used to create radially or azimuthally polarized beams, which can be decomposed into two beams with right- and left-handed circular polarization carrying different orbital angular momentum (OAM or $\ell$ ). By using a quarter-waveplate, the two beams can be projected to orthogonal linear polarizations, so that scalar beams can be selected with the desired OAM [44].

In the experiments we employed two different s-plates: SP1 and SP2. Firstly, SP1 introduces a $\pi$-dephase (half wave) and a spatially varying fast axis ($\Phi_{\text{fast}} = \Phi/2$, where $\Phi$ is the azimuthal coordinate), which means that incident horizontal linear polarization can be converted into radial polarization. Then, with SP1 we can generate $\ell = \pm 1$ optical vortices as described in [44]. Secondly, SP2 has a $\pi/_2$-dephase (quarter wave) with azimuthally varying fast axis ($\Phi_{\text{fast}} = \Phi - \pi/4$), which translates incident circular polarization into a radially polarized vortex that can be used to produce optical vortices with $\ell = +2$ and $\ell = 0$ [44]. Furthermore, in Section 3 we will show results combining both plates to create additional OAMs.

In Fig. 3 we show the scheme for the generation of NIR and VIS optical vortices. The input beam (linear or circular polarization) propagates through the s-plate, producing the superposition of two circularly polarized beams with different OAM and opposite helicity. Then, a quarter-waveplate with its neutral axes at $\pm 45°$ with respect to a linear polarizer is used to project and select each individual OAM.

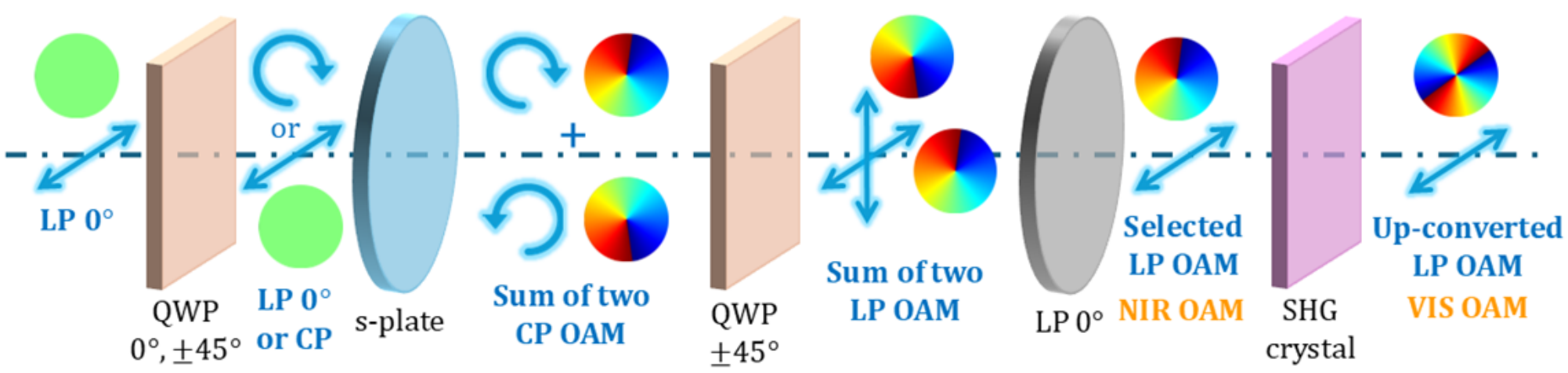

Figure 3. General scheme of the setup used for the generation of NIR and VIS vortices. The input beam is linearly polarized (LP). The first quarter-waveplate (QWP) keeps the LP or creates circular polarization (CP). The s-plates create different vortex vector beams. The second QWP decomposes the vortices to the desired orthogonal LP. The horizontal LP selects a vortex with the desired OAM in the NIR. A nonlinear crystal (BBO) converts the vortex to VIS through SHG process.

The NIR vortices were upconverted to the visible range (from 800 nm to 400 nm) using a type-I BBO crystal by means of second harmonic generation (SHG).

### *2.3. Temporal reference with the a-swing technique*

As discussed in Section 2.1, the BLASHI measurement provides a relative characterization of the spatio-spectral and spatiotemporal couplings. For the full beam dynamics, a temporal (or spectral) characterization of the beam at a particular point is needed. In this work, we use the amplitude swing (a-swing) technique [45,51], which has been proven to be robust and versatile, very recently being applied to few-cycle [52] and to vector pulses [53]. Here, we use a-swing to measure the reference pulse in the fundamental domain (800 nm). The a-swing technique can be applied in different spectral ranges [54], while for the case of pulses in 400 nm the SHG produces signal in a band that is not accessible in air, so an alternative will be described after.

Briefly, a-swing uses a compact setup with a rotating multiple-order waveplate (MWP) and a linear polarizer to create two delayed and amplitude-varying replicas (controlled by the plate orientation) of the unknown pulse (see Fig. 4a). The interfering pulses are up-converted (SHG) in a nonlinear (NL) crystal, while this signal is spectrally resolved as a function of the retarder plate orientation, resulting in a two-dimensional trace. This trace encodes the spectral phase of the pulse to be measured that can be reconstructed with a devoted numerical strategy. In our case, we use the differential evolution algorithm described in [54].

For every NIR vortex beam, we measured the pulse with a-swing at a coordinate $(x,y) = (0,-0.5)$ mm. Note that we chose to measure the beam away from its center to avoid the singularity. Here, we present an example of the temporal reference measurement for the beam with ℓ = +2 that we will upconvert afterwards. In Fig. 4 we show the experimental and retrieved a-swing traces, as well as the pulse reconstructed in the spectral and temporal domains. In this case, for that reference point the pulse has a pulse duration of 74.0 fs (for a Fourier limit of 60.7 fs). This reference pulse is used to connect the frequency-resolved wavefronts measured with BLASHI, whose results will be presented in the next section.

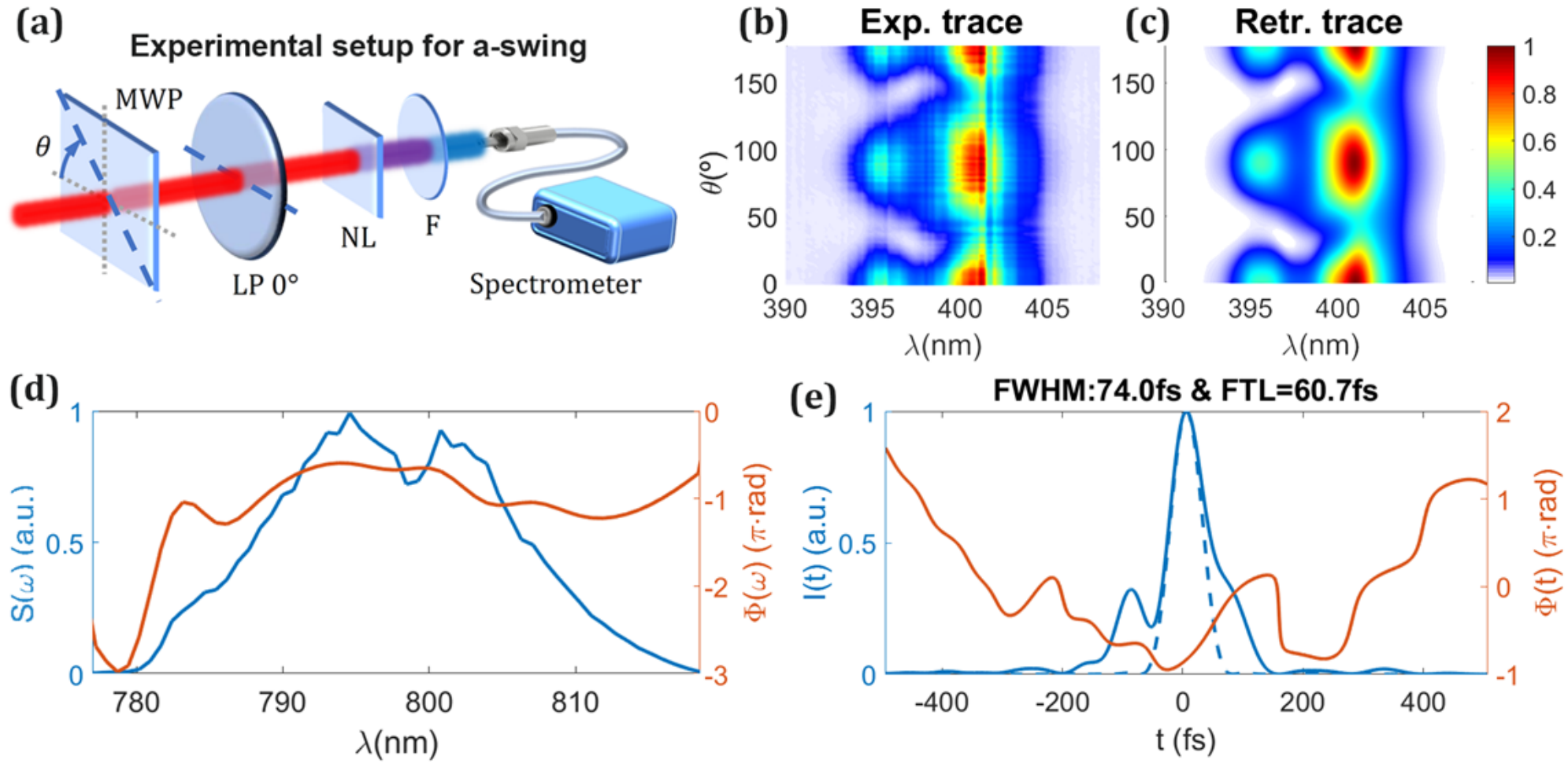

Figure 4. (a) Experimental setup for a-swing: rotating multiple-order waveplate (MWP), linear polarizer (LP), nonlinear crystal, filter (F) to remove the remaining fundamental signal before the spectrometer. (b) Experimental and (c) retrieved a-swing trace. (d) Spectral and (e) temporal intensity and phase of the measured pulse. Example corresponding to a NIR vortex with ℓ = +2.

Regarding the temporal measurement of the reference pulse at the VIS (400 nm), we do not use directly a-swing since we cannot detect the nonlinear signal at 200 nm (ultraviolet range). Here, we follow a different strategy that is to measure the reference pulse for the driving fundamental beam (800 nm) before creating the VIS vortex through SHG. Since we can model the SHG as the square of the pulse in the temporal domain (or the autoconvolution in the spectral domain), this reference pulse is translated to the VIS domain and is used to bridge the spatio-spectral phase obtained with BLASHI for the VIS vortices. This SHG model is valid due to the use of a thin nonlinear crystal, which ensures the assumptions of an undepleted pump and perfect phase-matching efficiency are met. The results will be presented in the next section.

## 3. Results and discussion

For the experiments, the light source employed was a Ti:sapphire chirped pulse amplification laser system (Spitfire ACE, Spectra Physics), delivering pulses with ~798 nm central wavelength, at a repetition rate of 5 kHz, Fourier-transform limit duration of ~60 fs and millijoule-level pulse energy. Using a 5 kHz laser system and a spectrometer with a minimum integration time of 2 ms resulted in the acquisition of multiple pulses at each scan point. In this case, averaging multiple shots does not affect the interference pattern due to the exceptional stability of the bulk interferometer, as discussed in the Introduction.

In the following sections, we first present the results of the characterization for the near-infrared optical vortices generated directly from the laser source. Then, we show the characterization of the visible optical vortices produced from SHG of the driving vortices, as described in Section 2.2. To have a complete spatiotemporal reconstruction, we will use in each case the measurement of a reference pulse with the a-swing technique.

### *3.1. Measurement of near-infrared driving optical vortices*

After demonstrating the technique, we employed it to analyze optical vortices with various OAMs, showcasing the method's effectiveness with different complex beams. We used a 1.06-mm-thick calcite walk-off crystal with optical axis oriented at $\theta_{OA} = 45°$ with respect to the normal to the input face, as described in Section 2.1. This is translated into a walk-off step of 113 μm at λ=800 nm.

In Section 2.2, we explained how to generate optical vortices with different OAMs based on the incident polarization and the specific s-plate used. Using SP1 and the combination of SP1+SP2, we created vortices with $\ell = \pm1$ and $\ell = +3$. These were measured with 113 μm (walk-off) scanning steps, and the results for the central wavelength (λ = 798 nm) are summarized in Fig. 5. The reference pulse at a position $(x,y) = (0, -0.5)$ mm was measured with a-swing (see Section 2.3).

Complete spatio-spectral retrievals of each optical vortex are available in Visualization 1 showing the $\ell = -1$, $\ell = +1$, and $\ell = +3$ near-infrared optical vortices, respectively. The spatio-spectral amplitude and phase representation (for the full spectral bandwidth) is equivalent to the spatiotemporal amplitude and phase evolution just by Fourier-transforming. Note that the s-plates are designed to operate in a bandwidth that covers the full spectrum of the fundamental pulse (~25 nm, centered at 800 nm), so that their response is achromatic in said bandwidth. This explains that the intensity and phase of the beam profile has the same pattern for the different wavelengths with the pulse bandwidth.

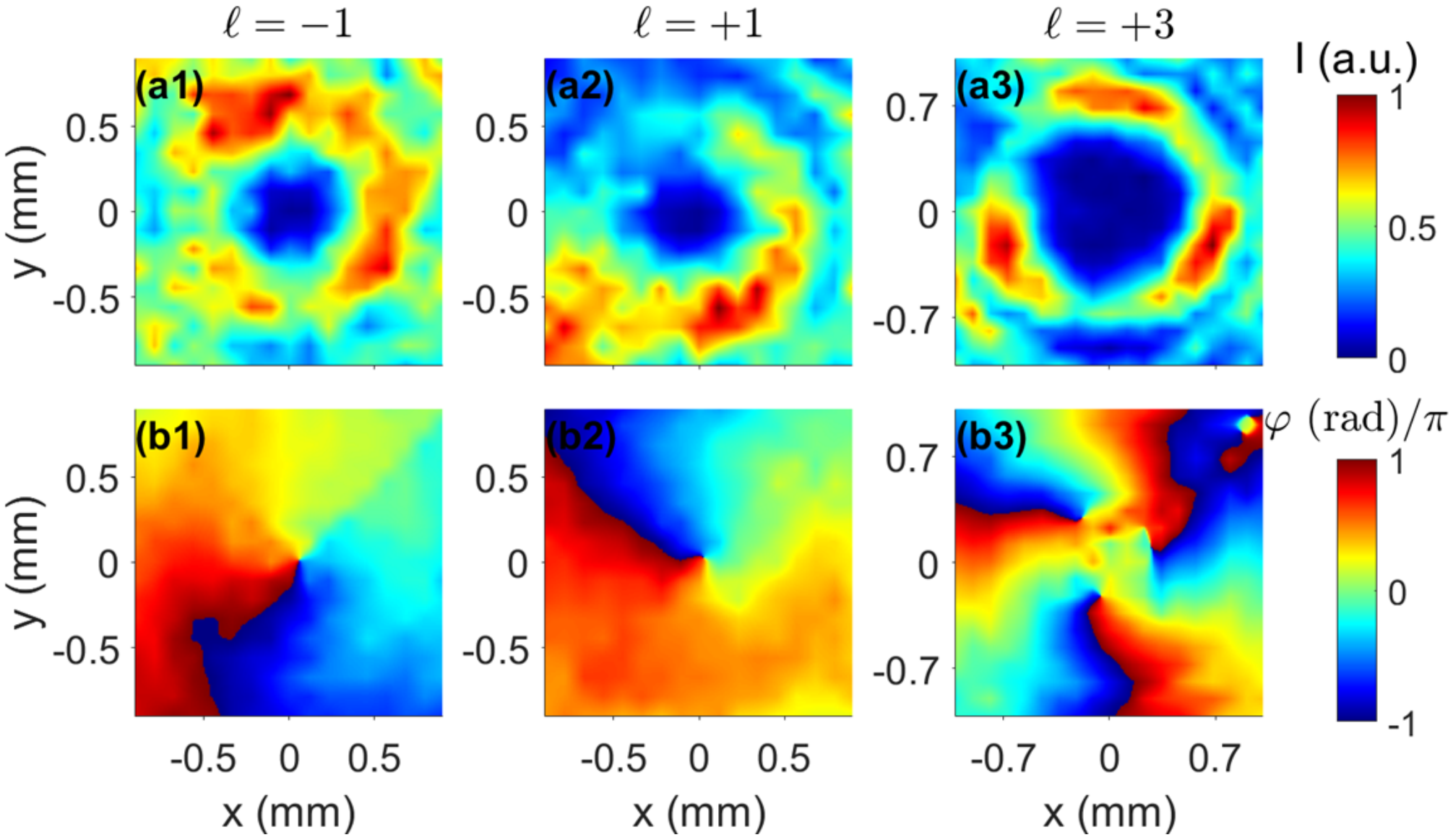

Figure 5. Intensity (a) and phase (b) obtained from the spatio-spectral characterization with BLASHI for the central wavelength (λ = 798 nm) of three different optical vortices with OAMs: $\ell = -1$ (column 1), $\ell = +1$ (column 2) and $\ell = +3$ (column 3). The results for the full spectral range are shown in Visualization 1. Note: the retrievals are shown interpolated with a spatial step of 10 μm.

As seen in Fig. 5, the spatial intensity profiles of these vortices are somewhat distorted due to imperfect generation that may be attributed to diffraction at the singularity of the s-plates. Note that the s-plates are fabricated by inscribing microstructures with varying orientations based on the azimuthal coordinate. However, at the center of the s-plate, the orientation is undefined—hence the singularity. This local feature can account for the observed distortions due to diffraction during propagation. In our setup, free-space propagation occurs after the s-plate, with a propagation distance of approximately 15-20 cm to the device. Despite these intensity distortions, the phase profiles remain clear, accurately reflecting each vortex OAM. Similar distortions were observed with the same s-plates in previous works [31,44], where it was discussed that they are likely associated with the use of SP1. It is important to note that in those studies, a different technique (STARFISH [26]) was used to measure the beam. Additionally, the spatially resolved spectrum (amplitude only) provides a direct measurement of the beam profile by scanning with the fiber spectrometer. As such, the observed profile structure is a direct measurement and cannot be attributed to detection artifacts or reconstruction errors. Therefore, we attribute these beam profiles to the inherent structure of the field generated by the s-plates.

### *3.2. Measurement of visible optical vortices*

Beyond its compactness and stability, a notable advantage of the spatiotemporal technique BLASHI is its applicability across the transparency range of the uniaxial birefringent crystal. The only requirement for operating in a different spectral range is recalculating the walk-off for the new region and having the appropriate standard spectrometer. As detailed in Section 2.1, for the 1.06-mm-thick calcite crystal ($\theta_{\mathrm{OA}} = 45°$), the walk-off at 400 nm is 123 μm, slightly larger than at λ = 800 nm.

Since the s-plates employed in this work are designed for 800 nm, we cannot directly generate VIS optical vortices with a driving VIS beam. As explained in Section 2.2, we opted to produce the VIS vortices by frequency-doubling the 800-nm vortices using a SHG crystal. We used a SHG BBO to transform the ℓ = ±1, +2 NIR-vortices into VIS vortices. To achieve complete spatio-spectral or spatiotemporal characterization, the temporal reference of the visible vortices at a single spatial position, $(x,y) = (0,-0.5)$ mm, was obtained from the a-swing measurement of the fundamental driving field as described in Section 2.3. The resulting reference VIS pulses in the spectral domain are shown in Fig. 6 for the VIS optical vortices with ℓ = –2, +2 and +4, respectively.

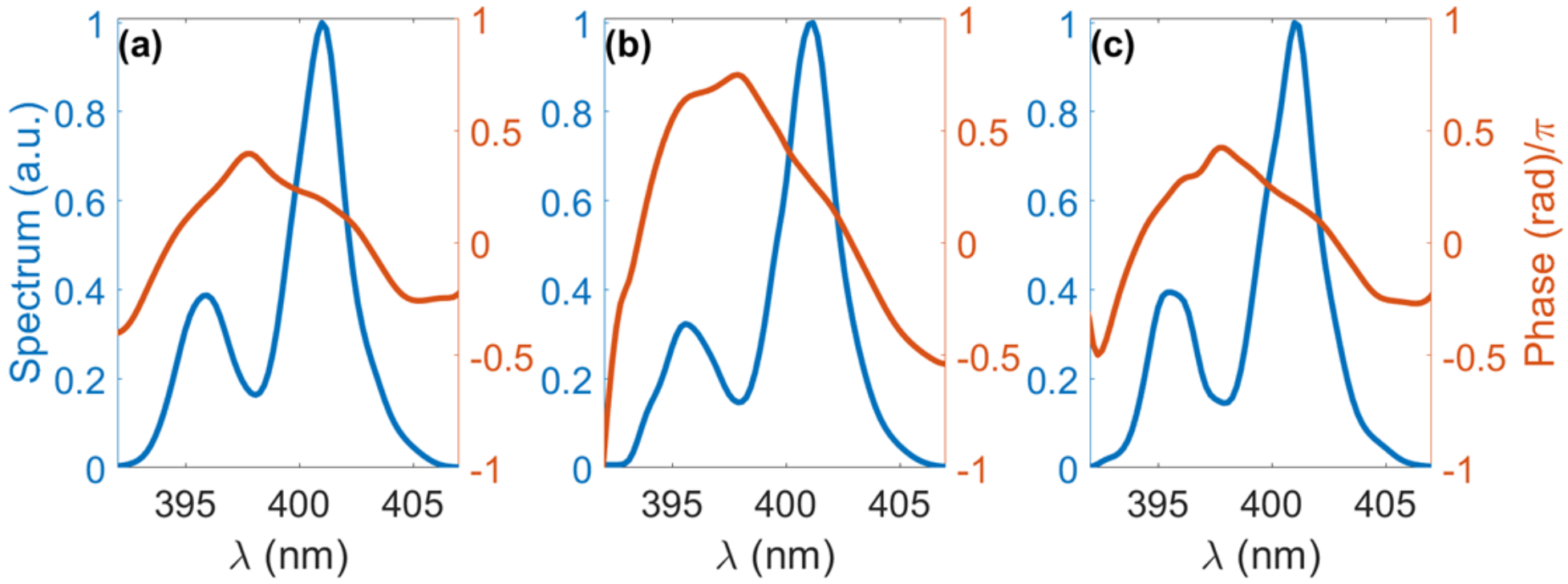

Figure 6. Spectral reference (intensity and phase) measurement for the visible vortices simulated from the optical vortices in the NIR range, obtained from the a-swing measurement of the fundamental pulse, for the VIS vortices with OAMs: (a) ℓ = −2, (b) ℓ = +2, and (c) ℓ = +4.

Knowing the spectral references and measuring the spatio-spectral gradients with the BLASHI technique allows us to fully characterize the visible vortices. The spatio-spectral retrieval for the central wavelength (λ = 396.7 nm) for each vortex is shown in Fig. 7, with complete retrievals in Visualization 2, for the visible optical vortices with ℓ = −2, ℓ = 2, and ℓ = 4.

The spatio-spectral phase profiles (second row of Fig. 7) show that the OAM has doubled, in agreement with the angular momentum conservation [55] that is basically a nonlinear scaling rule. This rule states that the OAM of the q-th order harmonic generation process is $\ell_q = q \cdot \ell_0$, where $\ell_0$ is the fundamental frequency's OAM. Here, we observe doubled OAMs because the nonlinear order is q = 2 (SHG).

Conversely, the spatio-spectral intensity profile (first row of Fig. 7) reveals that the ℓ = ±2 vortices are asymmetrical and significantly distorted compared to the ℓ = +4 vortex. Comparing the visible and NIR vortices, those generated with SP1 are more distorted in the visible range, which is expected since SHG amplifies intensity distortions due to its quadratic scaling with fundamental intensity. Therefore, we obtain the same conclusion than for NIR vortices, that SP2 imprints better spatial distribution than SP1. As discussed at the end of Section 3.1, the structured profile originates from the singularity in the s-plate microstructures, and it is a direct measurement that cannot be attributed to detection artifacts.

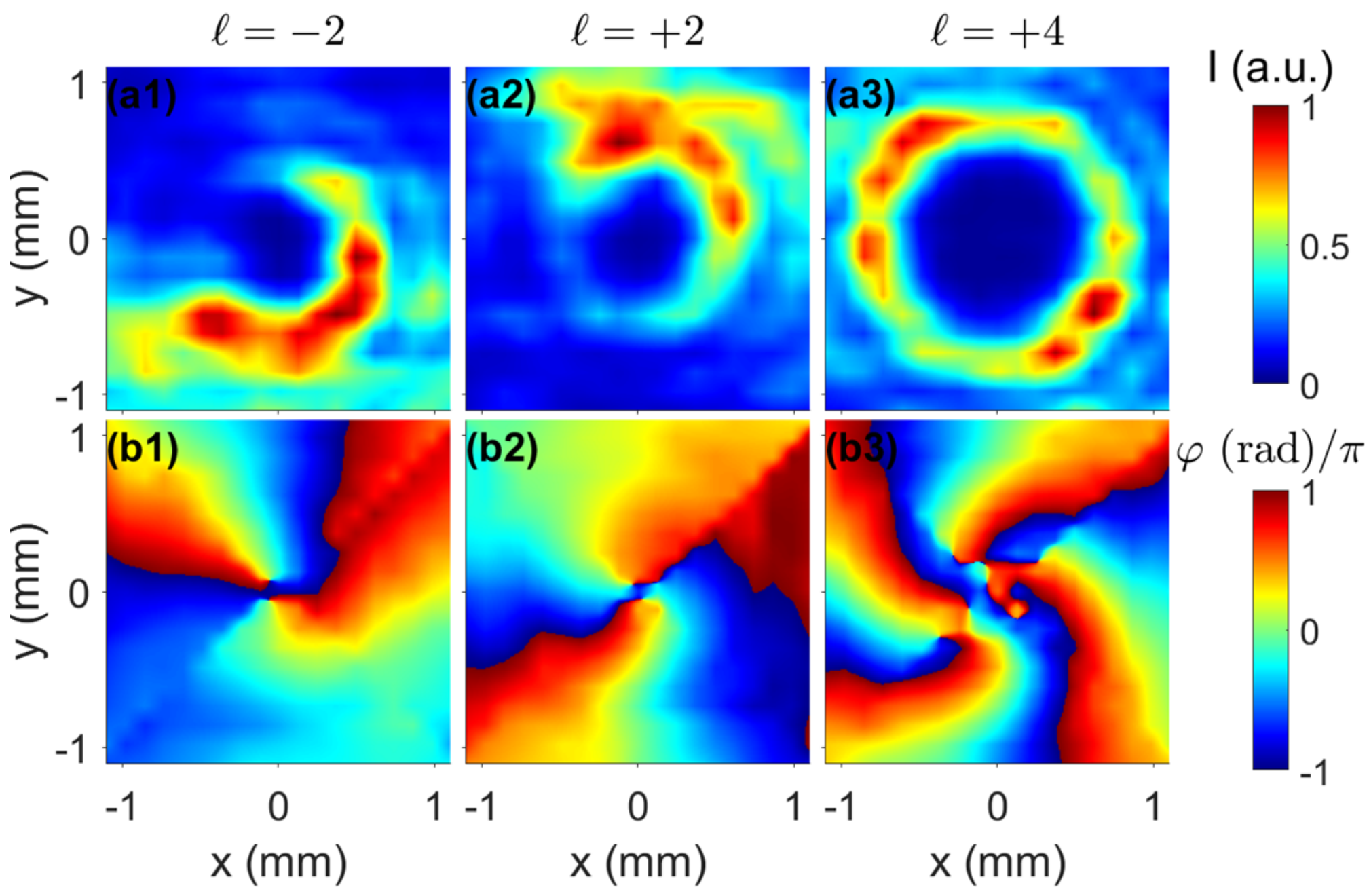

Figure 7. Intensity (a) and phase (b) from the spatio-spectral characterization with BLASHI for the central wavelength (λ = 396.7 nm) for optical vortices having different OAM within the visible spectrum: ℓ = −2 (first column), ℓ = +2 (second column) and ℓ = +4 (third column). The results for the full spectral range are shown in Visualization 2. Note: the retrievals are shown interpolated with a spatial step of 10 μm.

## 4. Conclusions

We have applied recently developed techniques to characterize ultrashort laser beams in different spectral regions. The BLASHI technique has been used to measure the spatio-spectral and spatiotemporal couplings, while the a-swing technique has been used to measure a reference pulse, required to have the absolute beam structure rather than the relative couplings. Both techniques are interferometric but use bulk configurations based on polarization shaping and control, so that they are compact, ultrastable and versatile.

Thanks to using simple elements as uniaxial crystals and linear polarizers, the BLASHI technique can operate in different spectral ranges with the same setup. Here, we have shown operation in the near-infrared and the visible range, while the operation range is wider in different bands across the spectrum whereas the uniaxial crystal is transparent. We have studied the spectral response of the detection system, finding that the spatial scanning step in the measurement needs to be adjusted to match the crystal walk-off at a particular spectral range.

The near-infrared optical vortices were generated in-line from vector beams produced by s-plates. Those vortices were used as driving fields to produce the visible vortices by nonlinear up-conversion. We have found consistent results with the theory of vector beam generation and manipulation. We have also discussed the range of operation of the s-plates and analyzed feature deviations from ideal behavior.

**Funding.** Ministerio de Ciencia e Innovación (PID2020-119818GB-I00, PID2023-149836NB-I00); Consejería de Educación, Junta de Castilla y León (SA108P24); European Regional Development Fund.

**Disclosures.** The authors declare no conflicts of interest.

**Data availability.** Data underlying the results presented in this paper are not publicly available at this time but may be obtained from the authors upon reasonable request.